\documentclass[10pt,twocolumn]{article}
\usepackage{graphicx}
\usepackage{amsmath} %non-Springer
\usepackage{amssymb} %non-Springer
\usepackage{fancyvrb} %non-Springer
\usepackage{soul}
\usepackage{fancyhdr}
\usepackage{hyperref}
\makeatletter
\renewcommand{\maketitle}{\bgroup
\begin{flushleft}
  \begin{Huge}
  \textbf{\@title}\\
  \end{Huge}
  \vspace{1cm}
  \@author
\end{flushleft}\egroup
}
\makeatother
\title{Detecting \textit{t\={a}la} Computationally in Polyphonic Context - A Novel Approach}
\author{%
    \textbf{{\large Susmita Bhaduri}}$^{1}$, \textbf{{\Large Anirban Bhaduri}}$^{2}$, \textbf{{\Large Dipak Ghosh}}$^{3}$\\
    $^{1,2,3}$Deepa Ghosh Research Foundation, Kolkata-700031,India \\
    \underline{$^{1}$susmita.sbhaduri@dgfoundation.in}\\
    \underline{$^{2}$bhaduri.anirban@dgfoundation.in}\\
    \underline{$^{3}$deegee111@gmail.com}
}

\begin{document}
\twocolumn[
  \begin{@twocolumnfalse}
    \maketitle
  \end{@twocolumnfalse}
  ]
\noindent

\begin{abstract}
In North-Indian-Music-System(NIMS),\textit{tabl\={a}} is mostly used as percussive accompaniment for vocal-music in polyphonic-compositions. The human auditory system uses perceptual grouping of musical-elements and easily filters the \textit{tabl\={a}} component, thereby decoding prominent rhythmic features like \textit{t\={a}la}, tempo from a polyphonic-composition. For Western music, lots of work have been reported for automated drum analysis of polyphonic-composition. However, attempts at computational analysis of \textit{t\={a}la} by separating the \textit{tabl\={a}}-signal from mixed signal in NIMS have not been successful. \textit{Tabl\={a}} is played with two components - right and left. The right-hand component has frequency overlap with voice and other instruments. So, \textit{t\={a}la} analysis of polyphonic-composition, by accurately separating the \textit{tabl\={a}}-signal from the mixture is a baffling task, therefore an area of challenge. In this work we propose a novel technique for successfully detecting \textit{t\={a}la} using left-\textit{tabl\={a}} signal, producing meaningful results because the left-\textit{tabl\={a}} normally doesn't have frequency overlap with voice and other instruments. North-Indian-rhythm follows complex cyclic pattern, against linear approach of Western-rhythm. We have exploited this cyclic property along with stressed and non-stressed methods of playing \textit{tabl\={a}}-strokes to extract a characteristic pattern from the left-\textit{tabl\={a}} strokes, which, after matching with the grammar of \textit{t\={a}la}-system, determines the \textit{t\={a}la} and tempo of the composition. 
A large number of polyphonic(vocal+\textit{tabl\={a}}+other-instruments) compositions has been analyzed with the methodology and the result clearly reveals the effectiveness of proposed techniques.
\end{abstract}


\section{Introduction}
\label{intro}
Current research in Music-Information-Retrieval(MIR) is largely limited to Western music cultures and it does not address the North-Indian-Music-System hereafter NIMS, cultures in general. NIMS raises a big challenge to current rhythm analysis techniques, with a significantly sophisticated rhythmic framework. We should consider a knowledge-based approach to create the computational model for NIMS rhythm. Tools developed for rhythm analysis can be useful in a lot of applications such as intelligent music archival, enhanced navigation through music collections, content based music retrieval, for an enriched and informed appreciation of the subtleties of music and for pedagogy. 
Most of these applications deal with music compositions of polyphonic kind in the context of blending of various signals arising from different sources. Apart from the singing voice, different instruments are also included. 

As per~\cite{london2004} rhythm relates to the \textit{patterns of duration} that are phenomenally present in the music. It should be noted that that these \textit{patterns of duration} are not based on the actual duration of each musical event but on the Inter Onset Interval(IOI) between the attack points of successive events. As per[~\cite{cooper1960}], an accent or a stimulus is marked for consciousness in some way. Accents may be phenomenal, i.e. changes in intensity or changes in register, timbre, duration, or simultaneous note density or structural like arrival or departure of a cadence which causes a note to be perceived as accented. It may be metrical accent which is perceived as accented due to its metrical position[~\cite{lerdahl1985}. Percussion instruments are normally used to create accents in the rhythmic composition. The percussion family which normally includes timpani, snare drum, bass drum, cymbals, triangle, is believed to include the oldest musical instruments, following the human voice[~\cite{latham2011}]. The rhythm information in music
is mainly and popularly provided by the percussion instruments. One simple way of analyzing rhythm of a composite or polyphonic music signal having some percussive component, may be to extract the percussive component from it using some source separation techniques based on frequency based filtering. Various attempts have been made in Western music to develop applications for re-synthesizing the drum track of a composite music signal, identification of type of drums played in the composite signal[ex. the works of~\cite{gillet2005,ono2008} etc., described in Section~\ref{past} in detail]. Human listeners are able to perceive individual sound events in complex compositions, even while listening to a polyphonic music recording, which might include unknown timbres or musical instruments. However designing an automated system for rhythm detection from a polyphonic music composition is very difficult. 

In the context of NIMS rhythm popularly known as \textit{t\={a}la}, \textit{tabl\={a}} is the most popular percussive instrument. Its right hand drum-\textit{dayan} and left hand drum-\textit{bayan} are played together and amplitude-peaks spaced at regular time intervals, are created by playing every stroke.
One way of rhythm information retrieval from polyphonic composition having \textit{tabl\={a}} as one of the percussive instruments, may be to extract the \textit{tabl\={a}} signal from it and analyze it separately. 
The \textit{dayan} has a frequency overlap with other instruments and mostly human-voice for polyphonic music, so if we extract the whole range of frequencies for both \textit{bayan} and \textit{dayan} components, by existing frequency based filtering methods, the resultant signal will be a noisy version of original song as it will still have  part of other instruments, human voice components along with \textit{tabl\={a}}. Also conventional source separation methods lead to substantial loss of information or sometimes addition of unwanted noise. This is the an area of challenge in \textit{t\={a}la} analysis for NIMS. Although, NIMS \textit{t\={a}la} functions in many ways like Western meter, as a periodic, hierarchic framework for rhythmic design, it is composed of a sequence of unequal time intervals and has longer time cycles. Moreover \textit{t\={a}la} in NIMS is distinctively cyclical and much more complex compared to Western meter[~\cite{clayton1997}]. This complexity is another challenge for \textit{t\={a}la} analysis.
 
Due to the above reasons defining a computational framework for automatic rhythm information retrieval for North Indian polyphonic compositions is a challenging task. Very less work has been done for rhythmic information retrieval from a polyphonic composition in NIMS context. In Western music, quite a few approaches are followed for this purpose, mostly in the areas of beat-tracking, tempo analysis, annotation of strokes/pulses from the separated percussive signal. We have described these systems in the Section~\ref{past}. For NIMS, very few works of rhythm analysis are done by adopting Western drum-event retrieval system. These works result in finding out meter or speed which are not very significant 
in the context of NIMS. Hence this is an unexplored area of research for NIMS.

In this work we have proposed a completely new approach, i.e.instead of extracting both \textit{bayan} and \textit{dayan} signal, we have extracted the \textit{bayan} signal from the polyphonic composition by using band-pass filter. This filter extracts lower frequency part which normally does not overlap with the frequency of human voice and other instruments in a polyphonic composition. Most of the \textit{t\={a}la}-s start with a \textit{bol} or stroke which has a \textit{bayan} component(either played with \textit{bayan} alone or both \textit{bayan} and \textit{dayan} together) and also the some consequent section(\textit{vibh\={a}ga} in NIMS terminology) boundary-\textit{bol}-s have similar \textit{bayan} component. Hence these strokes would be captured in the extracted \textit{bayan} signal.
For a polyphonic composition, its \textit{t\={a}la} is rendered with cyclically recurring patterns of fixed time-lengths. This is the cyclic property of NIMS, discussed in detail in section~\ref{defn}.
So after extracting the starting \textit{bol}-s and the section boundary strokes from the \textit{bayan} signal, we can exploit the cyclic property of a \textit{t\={a}la} and the pattern of strokes appearing in a single cycle and can detect important rhythm information from a polyphonic composition.
This would be a positive step towards rhythm information retrieval from huge collection of
music recordings for both film music and live performances of various genres of \textit{hindi} music.
Here, we consider the \textit{t\={a}la} detection of different single-channel, polyphonic clips of \textit{hindi} vocal songs of devotional, semi-classical and movie soundtracks from NIMS, having variety of tempo and 
\textit{m\={a}tr\={a}}-s.

The rest of the paper is organized as follows. A review of past work is presented
in section~\ref{past}. Some definitions are provided in section~\ref{defn}. In section~\ref{meth} the
proposed methodology is elaborated. Experimental results are placed in section~\ref{exp} and the paper ends with concluding remarks in section~\ref{con}. 

\section{Definitions}
\label{defn}
\subsection{\textbf{\textit{T\={a}la} and its structure in NIMS}}
\label{tala}

\begin{figure}[!t]
\centering
\includegraphics[width=0.5\textwidth]{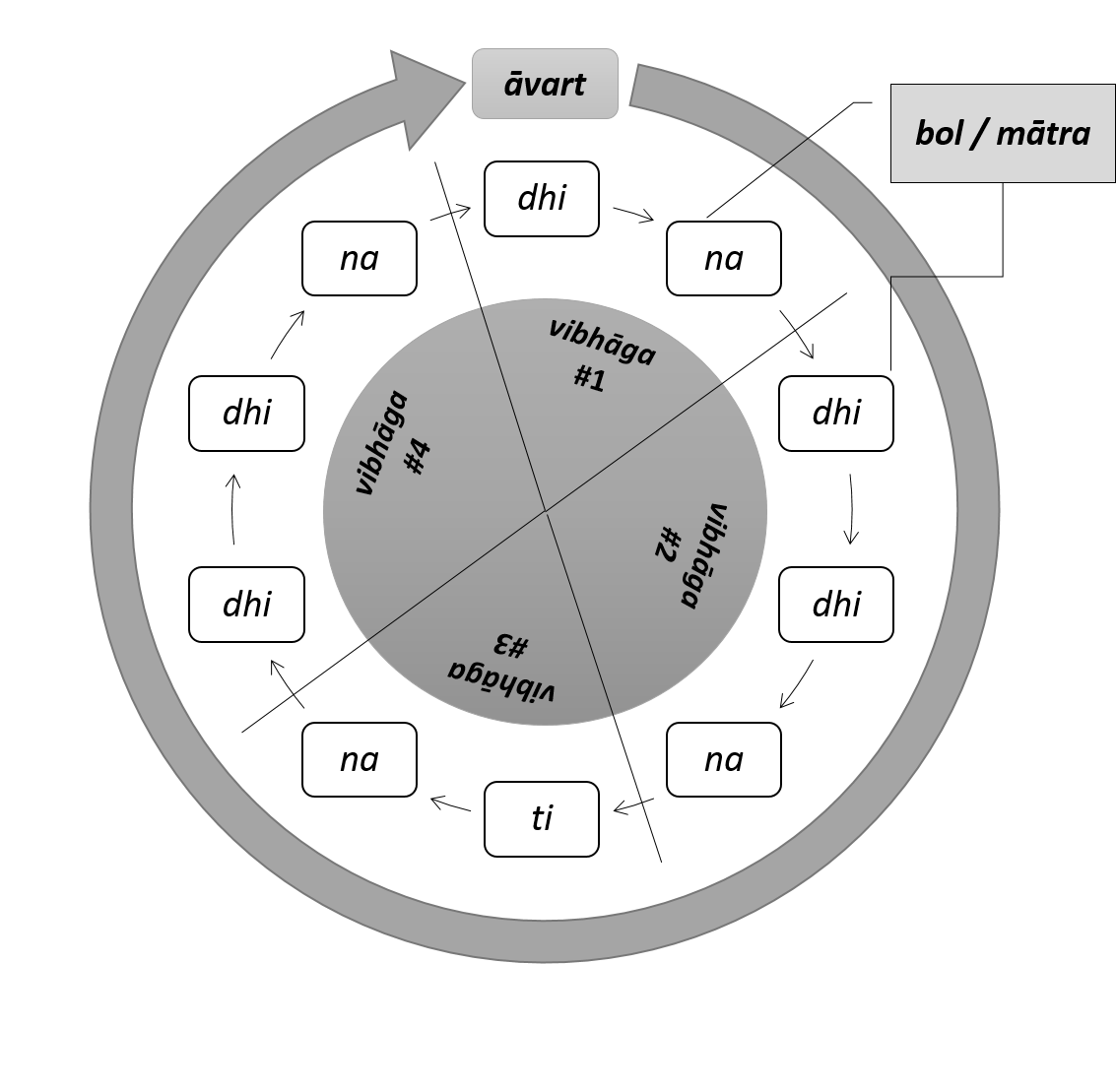}
\caption{The hierarchy of NIMS \textit{t\={a}la} illustrated with \textit{jhaptal}}
\label{heir}
\end{figure}

The basic identifying features of rhythm or \textit{t\={a}la} in NIMS are described as follows. 
\begin{itemize}
\item \textbf{\textit{T\={a}la} and its cyclicity: } North Indian music is metrically organized and it is called \textit{nibaddh}(bound by rhythm) music. This kind of music is set to a metric framework called \textit{t\={a}la}.
Each \textit{t\={a}la} is uniquely represented as cyclically recurring patterns of fixed time-lengths. 

\item \textbf{\textit{\={A}vart}: }This recurring cycle of time-lengths in a \textit{t\={a}la} is called \textit{\={a}vart}. \textit{\={A}vart} is used to specify the number of cycles played in a composition, while annotating the composition.

\item \textbf{\textit{M\={a}tra: }}The overall time-span of each cycle or \textit{\={a}vart} is made up of a certain number of smaller time units called \textit{m\={a}tra}-s. The number of \textit{m\={a}tra}-s for the NIMS \textit{t\={a}la}-s, usually varies from $6$ to $16$. 

\item \textbf{\textit{Vibh\={a}ga:} }The \textit{m\={a}tra}-s of a \textit{t\={a}la} are grouped into sections, sometimes with unequal time-spans, called \textit{vibh\={a}ga}-s.
%\textit{Vibh\={a}ga}-s are indicated through the hand gestures of a \textit{t\={a}l\={i}}(clap or \textit{sasabda kriya}) and a \textit{khali}(wave or \textit{nisabda kriya}), as explained below.

\begin{figure}
\centering
\includegraphics[width=0.5\textwidth]{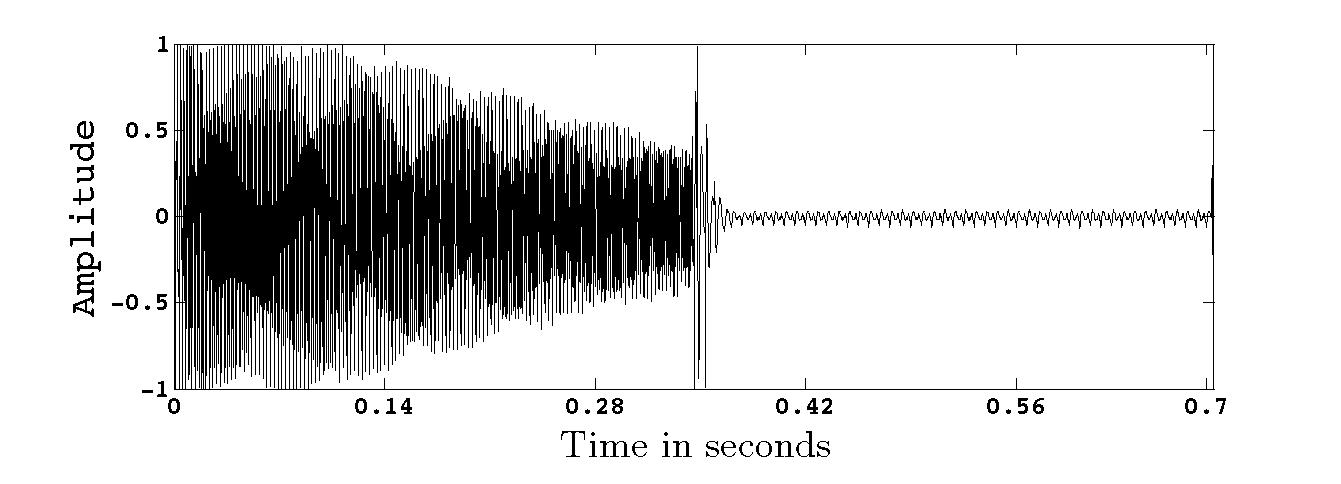}
\caption{\textit{ta}*(absent-\textit{bol})}
\label{absent}
\end{figure}

\item \textbf{\textit{Bol: }}In the \textit{t\={a}la} system of North Indian music, the actual articulation of \textit{t\={a}la} is done by certain syllables which are the mnemonic names of different strokes/pulses corresponding to
each \textit{m\={a}tra}. These syllables are called \textit{bol}-s. There are four types of \textit{bol}-s as defined below.

\begin{enumerate}
\item\textbf{\textit{Sam:} }The first \textit{m\={a}tr\={a}} of an \textit{\={a}vart} is referred as \textit{sam} which is mandatorily stressed[~\cite{clay2000}]. 

\item\textbf{\textit{T\={a}l\={i}-bol}: }\textit{T\={a}l\={i}-bol}-s are usually stressed, whereas \textit{khali}-s are not. \textit{T\={a}l\={i}-bol}-s are gestured by the \textit{tabl\={a}} player with claps of the hands, hence are called \textit{sasabda kriya}. The \textit{sam} is almost always a \textit{T\={a}l\={i}-bol} for most of the \textit{t\={a}la}-s, with only exception of \textit{rupak} \textit{t\={a}la} which designates the \textit{sam} with a moderately stressed \textit{bol} called \textit{khali}(as explained below)[~\cite{david2013}]. 

Highly stressed \textit{vibh\={a}ga} boundaries are indicated through the \textit{t\={a}l\={i}-bol}s[\cite{david2013}]. \textit{T\={a}l\={i}-sam} is indicated with a ($+$) in the rhythm notation of NIMS. Consequent \textit{T\={a}l\={i}-vibh\={a}ga}-boundaries are indicated with $2,3,\ldots$.

\item\textbf{\textit{Khali-bol}: }\textit{Khali} literally means empty and for NIMS it implies wave of the hand or \textit{nisabda kriya}. Moderately stressed \textit{Vibh\={a}ga} boundaries are indicated through the \textit{khali-bol}s so we almost never find the \textit{khali} applied to strongly stressed \textit{bol}-s like \textit{sam}[~\cite{david2013}].

\textit{khali-sam} is indicated with a ($0$) in the rhythm notation of NIMS and consequent \textit{khali-vibh\={a}ga}-boundaries are indicated also with $0$.

\item\textbf{Absent-\textit{bol}: }Sometimes while playing \textit{tabl\={a}}, certain \textit{bol}-s are dropped maintaining the perception of rhythm intact. They are called rests and they have equal duration as a \textit{bol}. We have termed them as absent strokes/\textit{bol}-s. These \textit{bol}-s are denoted by $*$ in the rhythm notation of a NIMS composition~\href{http://www.ancient-future.com/pronuind.html}{Ancient-future}. In the Figure~\ref{absent}, the waveform of absent \textit{bol}, denoted by $*$, is shown just after another \textit{bol ta}, played in a \textit{tabl\={a}}-solo.

Normally in a NIMS composition there may be many absent \textit{bol}-s in the \textit{thek\={a}} played for the \textit{t\={a}la}. In these cases other percussive instruments(other than \textit{tabl\={a}}) and vocal emphasis might generate percussive peaks for the time positions of the absent strokes, depending on the composition, the lyrics being sung and thus the rhythm of the composition is maintained. 
\end{enumerate}

\item \textbf{\textit{Thek\={a}}: }For \textit{tabl\={a}}, the basic characteristic pattern of \textit{bol}-s that repeats itself cyclically along the progression of the rendering of \textit{t\={a}la} in a composition, is called \textit{thek\={a}}. In other words it's the most basic cyclic form of the \textit{t\={a}la}[~\cite{david2013}]. Naturally \textit{thek\={a}} corresponds to the basic pattern of \textit{bol}-s in an \textit{\={a}vart}.
%The \textit{thek\={a}} of \textit{t\={a}la}, is cyclically repeated over the entire length of the composition.
The strong starting \textit{bol} or \textit{sam} along with the \textit{t\={a}l\={i}}-\textit{vibh\={a}ga}-boundaries in a \textit{thek\={a}} carries the main accent and creates the sensation of cadence and cyclicity. 
%The correspondence between \textit{thek\={a}} and \textit{vibh\={a}ga} is shown in the Figure~\ref{hier}.
\end{itemize}

\textbf{Description of the definitions with an example:}

\begin{figure*}[!t]
\centering
\includegraphics[width=1.0\textwidth]{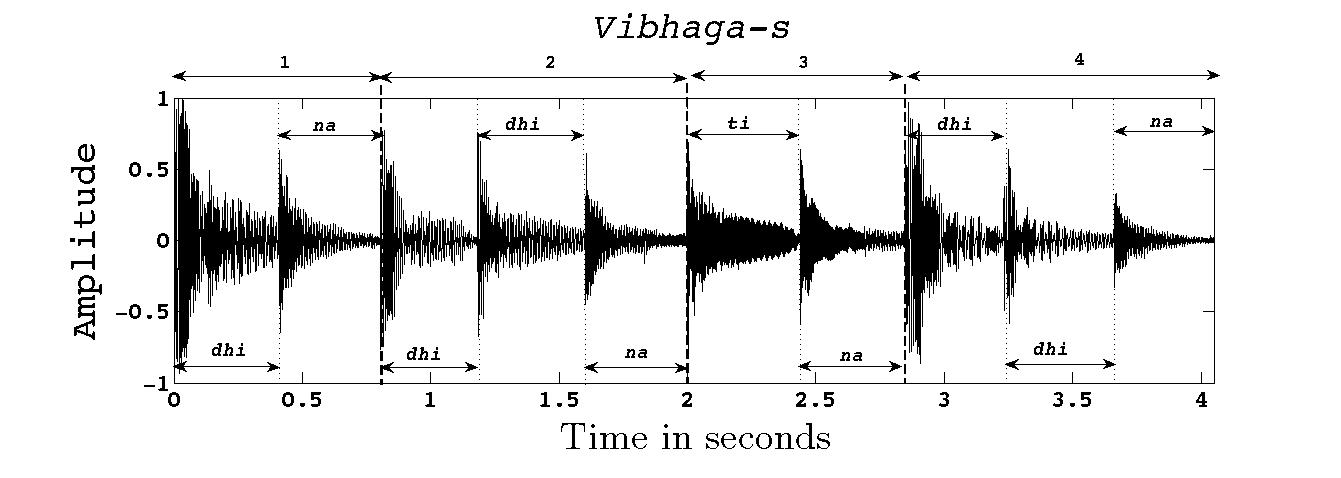}
\caption{The basic pattern of \textit{bol}-s for a single \textit{\={a}vart} or \textit{thek\={a}} of \textit{jhaptal}}
\label{fig_jhap}
\end{figure*}

 The details of these theories are shown in the structure of a \textit{t\={a}la}, called \textit{jhaptal} in the Table~\ref{jhap} and Figure~\ref{fig_jhap}. The hierarchy of the features and their interdependence are shown in the Figure~\ref{heir}. The cyclic property of \textit{t\={a}la} is evident here.

\begin{table*}
\caption{Description of \textit{jhaptal}, showing the structure and the its basic \textit{bol}-pattern or the \textit{thek\={a}}}
\label{jhap}
\begin{center}
%\begin{tabular}[width=0.75\textwidth]{|l|l|l|l|l|l|l|l|l|l|l|} 
\begin{tabular}[width=0.75\textwidth]{|l|l|l|l|l|l|l|l|l|l|l|}
\hline
\textit{t\={a}l\={i}}&+&  & 2& & &0 & & 3& &\\ 
\hline
\textit{bol}&\textit{dhi}&\textit{na}&\textit{dhi}&\textit{dhi}&\textit{na}&\textit{ti}&\textit{na}& \textit{dhi}&\textit{dhi}&\textit{na}\\ 
\hline
\textit{m\={a}tr\={a}}&1&2& 3& 4& 5& 6& 7& 8& 9& 10\\ 
\hline
\textit{vibh\={a}ga}&1&& 2& & & 3& & 4& & \\
\hline
\textit{\={a}vart}&1&& & & & & & & & \\
\hline
\end{tabular} 
\end{center}
\end{table*}

\begin{enumerate}
\item The first row of Table~\ref{jhap} shows the sequence of \textit{t\={a}l\={i}}-s and the \textit{khali}-s in a \textit{thek\={a}} or \textit{\={a}vart} of \textit{jhaptal}. In this row the \textit{sam} is indicated with a ($+$) sign and it should be noted that for \textit{jhaptal} it is a \textit{t\={a}l\={i}-bol}. 
The second \textit{t\={a}l\={i}}-\textit{vibh\={a}ga}-boundary is denoted by ($2$) followed by a ($0$) as it is the first \textit{khali}-\textit{vibh\={a}ga}-boundary and then by one more \textit{t\={a}l\={i}} denoted by ($3$) in a single cycle or \textit{thek\={a}}. 

The amplitude waveform of the same \textit{thek\={a}} is shown in the Figure~\ref{fig_jhap}. The \textit{sam} is shown in the Figure as the first \textit{bol dhi}. This clip of \textit{jhaptal} is available in~\href{http://www.tablaradio.com}{Tabla Radio}.

%\textit{Khali}-s are always indicated with $0$ in the notation. 
%Apart from \textit{sam}, the \textit{t\={a}l\={i}}-s, the \textit{khali}-s, other \textit{bol} positions are denoted by ($-$) in the first row of Table ~\ref{jhap}.

\item The second row of the Table~\ref{jhap} shows the \textit{bol}-s of \textit{jhaptal} in its \textit{thek\={a}}. 
In the Figure~\ref{fig_jhap} the waveform of all these \textit{bol}-s of a single cycle of \textit{jhaptal} are shown.

\item \textit{Jhaptal} \textit{thek\={a}} comprises of ten \textit{m\={a}tr\={a}}-s which are shown as per their sequence in third row of Table~\ref{jhap}. 

\item In the fourth row of the Table~\ref{jhap}, the section or \textit{vibh\={a}ga}-boundary-positions and sequences are shown. These \textit{vibh\={a}ga}-sequences are shown for \textit{jhaptal} in the Figure~\ref{fig_jhap}. We can see that there are four \textit{vibh\={a}ga}-s in \textit{jhaptal} \textit{thek\={a}} and first \textit{vibh\={a}ga}-boundary is a \textit{t\={a}l\={i}-sam-bol} \textit{dhi} having \textit{m\={a}tr\={a}} number as one. Second \textit{vibh\={a}ga}-boundary is again \textit{t\={a}l\={i}-bol} \textit{dhi} having \textit{m\={a}tr\={a}} number as three and so on.

\item In the fourth row of the Table~\ref{jhap}, \textit{\={a}vart}-position and sequence is shown. As there is one cycle shown so \textit{\={a}vart}-sequence is $1$.
\end{enumerate}

\subsection{\textbf{\textit{Tabl\={a}} and \textit{bol}-s}}
\label{tablaa}
\begin{figure}[!htb]
\minipage{0.5\textwidth}
  \includegraphics[width=\linewidth]{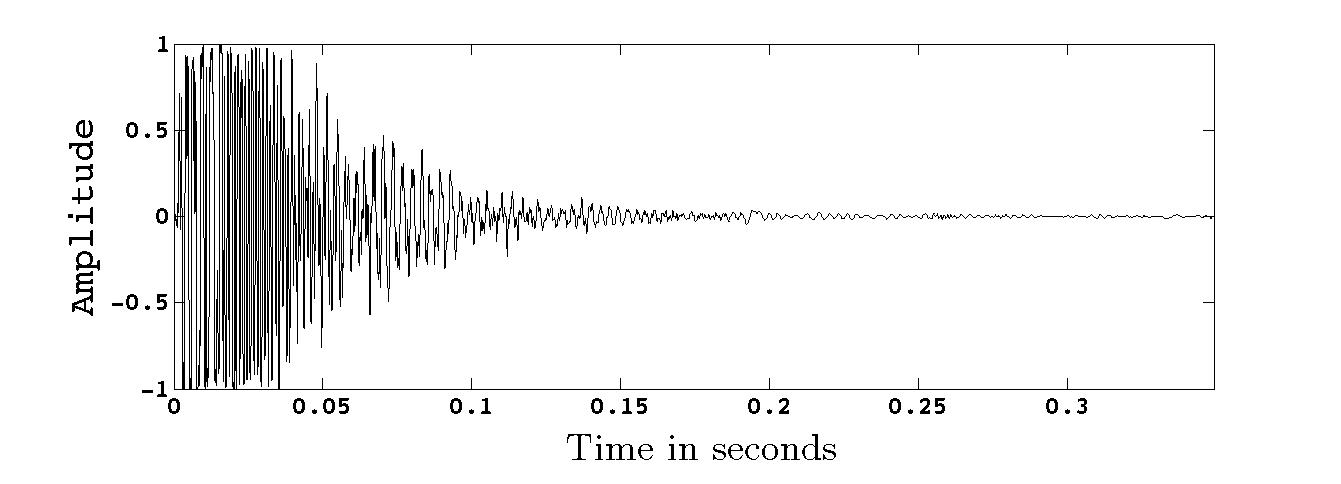}
  \caption{\textit{te(dayan)-bol}}\label{fig:te}
\endminipage\hfill
\minipage{0.5\textwidth}%
  \includegraphics[width=\linewidth]{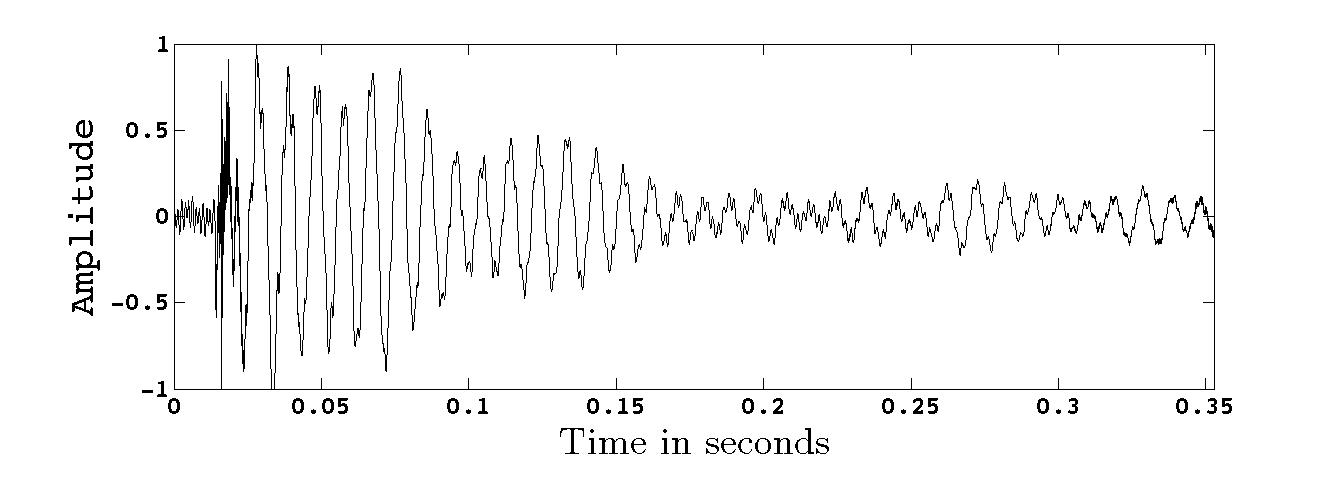}
  \caption{\textit{dha(bayan+dayan)-bol}}\label{fig:dha}
\endminipage
\end{figure}

\begin{figure}
\centering
\includegraphics[width=0.5\textwidth]{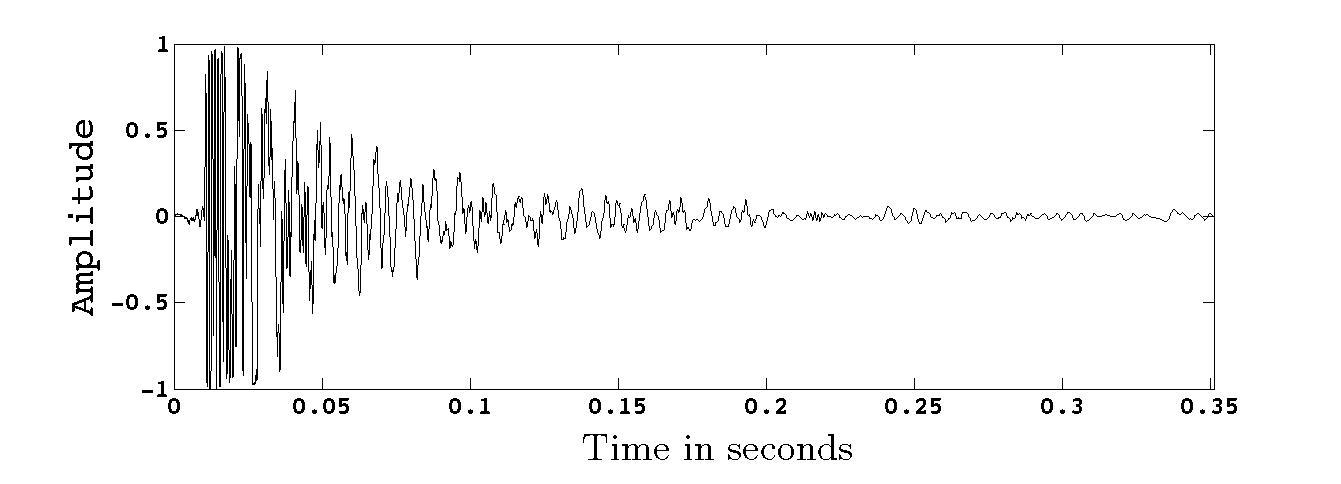}
\caption{\textit{ge(bayan)-bol}}
\label{ge}
\end{figure}

\textit{Tabl\={a}}, the traditional percussive accompaniment of NIMS, consists of a pair of drums. \textit{Bayan} the left drum, is played by the left hand and made with metal or clay. It produces loud resonant 
or damped non-resonant sound. As \textit{bayan} can not be tuned significantly, when it is played, it produces a fixed range of frequencies. The \textit{dayan} is the wooden treble drum, played by the right hand. A larger
variety of acoustics is produced on this drum when tuned in different frequency ranges.
%#susmita#
In the \textit{t\={a}la} system of North Indian music, the representation of \textit{t\={a}la}
is done mainly by playing \textit{bol}-s on the \textit{tabl\={a}}. 
\textit{bol}-s as they are played in \textit{tabl\={a}} are listed in Table~\ref{bol}. Figure~~\ref{fig:te}, \ref{fig:dha} and \ref{ge} shows the waveform of few sample waveforms of the \textit{bol}-s \textit{te, dha and ge} respectively. The clip of the \textit{bol}-s are taken from~\href{http://www.tablaradio.com}{Tabla Radio}.

\begin{table*}[t]
\caption{List of commonly played \textit{bol}-s in either on \textit{bayan} or \textit{dayan} or together on both}
\label{bol}
\begin{center}
\begin{tabular}{|c|c|l|}\hline
played on \textbf{\textit{bayan}}&played on \textbf{\textit{dayan}}&played on both \textbf{\textit{bayan}} and \textbf{\textit{dayan}}\\\hline
\textit{ke, ge ,ghe ,kath}&\textit{na, tin, tun, ti, te ,ta, da}& \textit{dha (na + ge), dhin (tin + ge), }\\
&& \textit{dhun (tun + ge), dhi (ti + ge)}\\\hline
\end{tabular}
\end{center}
\end{table*}

Most of the \textit{t\={a}la}-s have \textit{t\={a}l\={i}}-\textit{sam} played either with \textit{bayan}
alone or with \textit{bayan} and \textit{dayan} played simultaneously[~\cite{david2013}]. Same thing happens for the \textit{t\={a}l\={i}} \textit{vibh\={a}ga} boundaries. Most of the North Indian classical, semi-classical, devotional
and popular songs are played as per the \textit{t\={a}la}-s in Table~\ref{theka_table}. The most commonly played 
\textit{thek\={a}}-s are shown in this Table, Ref.\href{https://tabalchi.wordpress.com}{Tabla Class}; \href{http://www.tarang-classical-indian-music.com}{TAALMALA-THE RHYTHM OF MUSIC}.
For our experiment, we have considered the \textit{thek\={a}}-s listed in the Table~\ref{theka_table} for the \textit{t\={a}la}-s \textit{dadra}, \textit{kaharba}, \textit{rupak} and \textit{bhajani}. For these \textit{thek\={a}}-s, the stressed \textit{bol}-s having a \textit{bayan} component is shown in \textbf{bold} and pipes in \textbf{bold} indicate \textit{vibh\={a}ga} boundary. 
%These \textit{t\={a}la}-s are of distinct \textit{m\={a}tr\={a}}-s.

%\begin{figure}
%\centering
%\includegraphics[width=0.5\textwidth]{ge.jpg}
%\caption{\textit{ge(bayan)-bol}}
%\label{ge}
%\end{figure}

\begin{table*}[t]
\centering
\caption{Table of popular \textit{thek\={a}}-s of North Indian rhythms}
\label{theka_table}
\begin{tabular}{p{0.09\linewidth}|p{0.21\linewidth}|p{0.6\linewidth}}
\hline
\textbf{\textit{t\={a}la}} & \textbf{Number of \textit{m\={a}tr\={a}}-s/\textit{vibh\={a}ga} in an \textit{\={a}vart}} &\textbf{\textit{thek\={a}}}\\
\hline
\textit{\textbf{dadra}} & 3 I 3 & \textit{\textbf{dha} dhi na}I\textit{na ti na}\\
\hline
\textit{\textbf{kaharba}} & 4 I 4 & \textit{\textbf{dha} ge na ti}I\textit{na ke dhi na}\\
\hline
\textit{\textbf{rupak}} & 3 I 2 I 2 & \textit{tun na tun na ti te}I\textit{\textbf{dhin} dhin dha dha}I \textit{\textbf{dhin} dhin dha dha}\\
\hline
\textit{\textbf{bhajani}} & 4 I 4 & \textit{\textbf{dhin} $*$ na \textbf{dhin} $*$ dhin na $*$}I\textit{tin $*$ na tin $*$ tin na $*$}\\
\hline
\textit{jhaptal} & 2 I 3 I 2 I 3 & \textit{\textbf{dhi} na}I\textit{\textbf{dhi} dhi na}I\textit{ti na}I\textit{\textbf{dhi} dhi na}\\
\hline
\textit{tintal} & 4 I 4 I 4 I 4 & \textit{\textbf{dha} dhin dhin dha}I\textit{\textbf{dha} dhin dhin dha}I\textit{na tin tin na}I\textit{te-Te dhin dhin dha}\\
\hline
\end{tabular}
\end{table*}

%https://tabalchi.wordpress.com/2010/11/01/bhajani-theka-laggi/
%http://www.tarang-classical-indian-music.com/tabla_links/beats_and_phrases_of_tabla_eng.htm - dadra, kaharba
%bhajani-1 https://www.youtube.com/watch?v=ASxEJ81lnBw, http://tablabeats.blogspot.in/2011/06/bhajan-taal-1.html - bhajani-2
%http://searchgurbani.com/taal/taal_rupak
It is evident from the Table that all the \textit{t\={a}la}-s except \textit{rupak}, start with a \textit{t\={a}l\={i}}-\textit{sam-bol} having both \textit{dayan} and \textit{bayan} component. 
Only \textit{rupak} starts with a \textit{khali}-\textit{sam} and its \textit{sam} does not contain any \textit{bayan} component.
\textit{Bhajani} \textit{t\={a}la}, is often played with a variation for \textit{bhajan, kirtan} or \textit{qawwali} songs [~\cite{tabla2005}], which makes the first(\textit{t\={a}l\={i}}-\textit{sam}) and fourth \textit{bol} as stressed. Although this fourth \textit{bol} is not a \textit{t\={a}l\={i}} \textit{vibh\={a}ga} boundary still it is rendered as stressed and it is to be noted here that this \textit{bol} is played with \textit{dayan} and \textit{bayan} together. We have considered this \textit{bhajani} \textit{thek\={a}} with first(\textit{t\={a}l\={i}}-\textit{sam}) and fourth \textit{bol} as stressed, because data in our experiment includes popular \textit{bhajan} or devotional compositions.
 
For the \textit{t\={a}la}-s \textit{dadra} and \textit{kaharba} the number of
pulses in the \textit{thek\={a}} and the \textit{m\={a}tr\={a}}-s are identical but for \textit{rupak} 
and \textit{bhajani} each \textit{m\={a}tr\={a}} is divided in to two equal duration \textit{bol}-s. In effect
\textit{rupak} is a $7$ \textit{m\={a}tr\={a}} \textit{t\={a}la} but has $14$ pulses or \textit{bol}-s and \textit{bhajani} is $8$ \textit{m\={a}tr\={a}} \textit{t\={a}la} but has $16$ pulses or \textit{bol}-s. \textit{Bhajani} \textit{thek\={a}} in the Table~\ref{theka_table} has half of its number of strokes as absent \textit{bol}-s or rests(denoted by $*$).

It should be noted that the standard \textit{thek\={a}} of \textit{rupak} is 
\textit{tin$|$tin$|$na}I\textit{\textbf{dhin}$|$na}I\textit{\textbf{dhin}$|$na}, but we have taken another \textit{thek\={a}} shown in the 
Table~\ref{theka_table} for our experiment, Ref.[~\href{http://searchgurbani.com/taal/}{Search Gurbani}]. This \textit{thek\={a}} in normally followed for the semi-classical soundtracks and popular \textit{hindi} songs. Moreover we got maximum number of annotated samples of polyphonic songs composed with this \textit{thek\={a}} for our validation process.

\subsection{\textbf{\textit{Lay} or tempo}}
\label{lay}
An important concept of rhythm in NIMS is \textit{lay}, which governs the
tempo or the rate of succession of \textit{t\={a}la}. The \textit{lay} or tempo in NIMS can vary among \textit{ati-vilambit}(very slow), \textit{vilambit}(slow), \textit{madhya}(medium), \textit{dr̥uta}(fast) to \textit{ati-dr̥uta}(very fast).
%\href{http://chandrakantha.com/articles/indian_music/laya.html}{LAY (LAYA) - THE TEMPO}]. 
Tempo is expressed in beats per minute or BPM.

\section{Past work}
\label{past}
Although various rhythm analysis activities have been done for Western music, not much significant work has been done in the context of NIMS. Although the rhythmic aspect of Western music
is much simpler in comparison to Indian one, to get the broad idea of the problem, our study includes the work on Western music. The extraction of percussive events from a polyphonic composition is an ongoing and challenging area of research.  We have discussed existing drum separation approaches in Western music in Section~\ref{meth}, as they are relevant to our methodology. The existing works in meter analysis and beat-tracking for Western music, are discussed in the following section. Then similar discussion is made on existing rhythm analysis works in Indian music.

\subsection{\textbf{Meter analysis in Western music}}

In Western music beats have sharp attacks, fast decays and are uniformly repeated while in Middle Eastern and Indian music beats have irregular shapes. This is due the fact that bass instruments in these cultures are different from what is used in Western bands. By examining the distribution of Western meters,~\cite{algho1999} found that they deviate from Gaussianity by a larger amount than non-western meters.

Works have been done by parsing MIDI data into rhythmic levels by \cite{rosen92}. But that can not deal real audio data. \cite{bakh1997} attempted to encode the musical texts, notes into sequence of numbers and $\pm$ signs. But it can be implemented only for the Western compositions for which the notation is available.

\cite{todd96} showed multi-scale mechanism for the visualization of rhythm as rhythmogram. The rhythmogram provides a representation to the structure of spoken words and poems used, which is very different from polyphonic music but the model is implemented on synthesized binary pattern, strong and weak pulses, not from actual music composition.

\cite{algho1999} analysed the beat and rhythm information with a binary tree or trellis tree parsing depending on the length of the pauses in the input polyphonic signal.
The approach relies on beat and rhythm information extracted from the raw data after low-pass filtering. It has been tested using music segments from various cultures.

\cite{foote2000}, described methods for automatically locating points of significant change in music or audio, by analysing local self-similarity. This approach uses the signal to model itself, and thus does not rely on particular acoustic cues nor requires training.

\cite{klapuri2006} describes a method of estimating the musical meter jointly at three
metrical levels of measure, beat and subdivision, which are referred to as \textit{measure}, \textit{tactus} and
\textit{tatum}, respectively. For the initial time-frequency analysis, a new technique is proposed which measures the degree of musical accent as a function of time at four different frequency ranges. This is followed by a bank of comb filter resonators which extracts features for estimating the periods and phases of
the three pulses.The features are processed by a probabilistic model which represents primitive musical knowledge and uses the low-level observations to perform joint estimation of the tatum, tactus, and measure pulses.

\cite{guo2003} addressed the problem of classifying polyphonic musical audio signals of Western music, by their meter, whether duple/triple. Their approach aimed to test the hypothesis that acoustic evidences for downbeats can be measured on
signal low-level features by focusing especially on their temporal recurrences.

\subsection{Beat-tracking in Western music}
In the work of [\cite{bhat91}], a beat tracking system is described. A global tempo is first estimated.
A transition cost function is constructed based on the global tempo. Then dynamic programming is used to find
the best-scoring set of beat times that reflect the tempo.

In [~\cite{goto1995}] a real-time beat tracking system is designed, that processes audio signals that contain sounds of various instruments. The main feature of this work is to make context-dependent decisions by leveraging musical knowledge represented as drum patterns.

In [\cite{scheirer96}], the envelope of the music signal is extracted at different frequency
bands. The envelope information is then used to extract and track the strokes/pulses. 
\\
\\
To classify percussive events embedded in continuous audio streams,~\cite{gou2001} relied on a method based on automatic adaptation of the analysis frame size to the smallest metrical pulse, called the \textit{tick}. 

\cite{dixon2007} has created a system named BeatRoot for automatic tracking and annotation of
strokes for a wide range of musical styles. \cite{davies2007} proposed a context dependent beat tracking method
which handles varying tempos by providing a two state model. The first state tracks the tempo changes,
then the second maintains contextual continuity within a single tempo hypothesis. \cite{bock2011} proposed a data driven approach for beat tracking using context-aware neural networks.

\subsection{\textbf{Rhythm analysis in Indian Music}}
%\subsubsection{\textbf{Rhythm analysis and synthesis}}

The concepts of \textit{t\={a}la} and its elements are briefed in Section~\ref{defn}. 
%The structure of a \textit{t\={a}la} is complex. The structure of Western meter is comparatively regular and simpler.
For Indian percussive systems, strokes are of irregular nature and mostly are not of same strength. In comparison with Western music, not much significant work in rhythm analysis in Indian music, has been reported so far. 

The system proposed in [\cite{alex2009}] uses Probabilistic Latent Component Analysis method to extract \textit{tabl\={a}} signals from polyphonic \textit{tabl\={a}} solo. Then each separated signal is re-synthesized in each layer and the music is regenerated in \textit{quida}(improvisation of \textit{tabl\={a}} performances) model. The work is restricted to
\textit{tabl\={a}} solo performances where the \textit{tabl\={a}} signal is the most significant component, and not for polyphonic compositions where \textit{tabl\={a}} is one of the percussive accompaniment. 

In [~\cite{gulati2011}] the work of [~\cite{schull2007}] is extended. The methodology for meter
detection in  Western music is applied for Indian music. A two-stage comb filter-based approach, originally
proposed for double/triple meter estimation, is extended to a septuple meter (such as $7/8$ time-signature).
But this model does not conform to the \textit{t\={a}la} system of Indian music.

[~\cite{miron2011}] explored various techniques for rhythm analysis based on the Indian percussive
instruments. An effort is made to extract the \textit{tabl\={a}} component from a polyphonic music by
estimating the onset candidates with respect to the annotated onsets. Various existing segmentation techniques
for annotating polyphonic \textit{tabl\={a}} compositions, were also tried. But the goal of automatic
detection of \textit{t\={a}la} in Indian music did not succeed.

Some work has been done to detect a few important parameters like \textit{m\={a}tr\={a}}, tempo by first using signal level properties and then using cyclic properties of \textit{t\={a}la}.
The work in[\cite{icacni2014}] for \textit{m\={a}tr\={a}} and tempo detection for NIMS \textit{t\={a}la}-s, is based on the extraction of beat patterns that get repeated in the signal. Such pattern is identified by processing the amplitude envelope of a music signal. \textit{M\={a}tr\={a}} and tempo are detected from the extracted beat pattern.
This work is extended to handle different renderings of \textit{thek\={a}}-s comprised of single and composite \textit{bol}-s. In this work[\cite{eait2014}] \textit{bol} duration histogram is plotted from the \textit{beat signal} and the highest occurring \textit{bol}-duration is taken as the actual \textit{bol}-duration of the input \textit{beat signal}. 
The above methodology has been tested on electronic \textit{tabl\={a}} signal. In case of the real \textit{tabl\={a}} signal it is impossible to maintain consistency in terms of the periodicity of the {\it bol}-s or \textit{beat}-s played by a human.
To resolve this issue the work is further extended and modified to handle real \textit{tabl\={a}} signal in[~\cite{exica2014}] and this comparison is carried out for the entire \textit{beat signal} and a weight-age or the probability of the experimental signal being played according to certain \textit{t\={a}la}-s of NIMS, is calculated. The \textit{m\={a}tr\={a}} of the \textit{t\={a}la} for which this weightage is maximum, is confirmed as the \textit{m\={a}tr\={a}} of the input signal. This methodology was tested with real-\textit{tabl\={a}}-solo performance recordings. In recent times experiments and analysis have been done with non-stationary, nonlinear aspects of NIMS in [\cite{curr2016};\cite{maity2015}].

It is evident from the study that rhythm analysis in NIMS, focusing on \textit{t\={a}la} rendered with \textit{tabl\={a}}, the most popular North Indian percussive instrument, is a wide area of research. 
In our work, an approach for rhythm analysis is proposed, which is built around  the theory of \textit{t\={a}la} in NIMS.
%which is not an adaptation of well tried methodologies for Western music. Rather it is 

\section{Proposed Methodology}
\label{meth}
As it has been already discussed that there is a frequency overlap between \textit{tabl\={a}}(\textit{bayan} and \textit{dayan}) with voice and other instruments in a polyphonic composition, accurate extraction of \textit{tabl\={a}} signal from the mixed signal by following the source separation techniques based on frequency based filtering[~\cite{uhle2003};~\cite{hel2005}], has not been very successful. Also these source separation methods lead to substantial loss of information or sometimes addition of unwanted noise. 
It has motivated us to look for an alternate approach.
Here we have adopted a four-step methodology which is detailed out in following sections.
\begin{enumerate}
\item 
First we have processed the polyphonic input signal by partially adopting a filter-based separation technique. In doing so we are able to separate out the \textit{bayan}-stroke-signal which would consist of the only \textit{bayan}-strokes and also the \textit{bayan}-components of \textit{bayan}+\textit{dayan}-strokes.
\item
Then we have processed the entire polyphonic signal and generated a \textit{peak}-signal, which comprises of all the emphasized \textit{peak}-s generated out of \textit{tabl\={a}} and other percussive instruments played in the polyphonic composition of a specific \textit{t\={a}la}. \textit{Peak}-signal would contain the \textit{peak}-s of \textit{bayan}-stroke-signal, and also the emphasized \textit{peak}-s of \textit{tabl\={a}}(i.e. only \textit{dayan}-strokes and \textit{bayan}+\textit{dayan}-strokes). If other percussive instruments played, then in addition to the above, the \textit{peak}-signal would also contain emphasized \textit{peak}-s generated out of them
\item Next we have refined the \textit{bayan}-stroke-signal and the \textit{peak}-signal.
\item Lastly we propose to generate a co-occurrence matrix from both kinds of signals and exploit domain specific information of \textit{tabl\={a}} and \textit{t\={a}la} theory, to detect the \textit{t\={a}la} and tempo of the input polyphonic signal.
\end{enumerate}

\begin{figure*}[!t]
\centering
\includegraphics[width=1.0\textwidth]{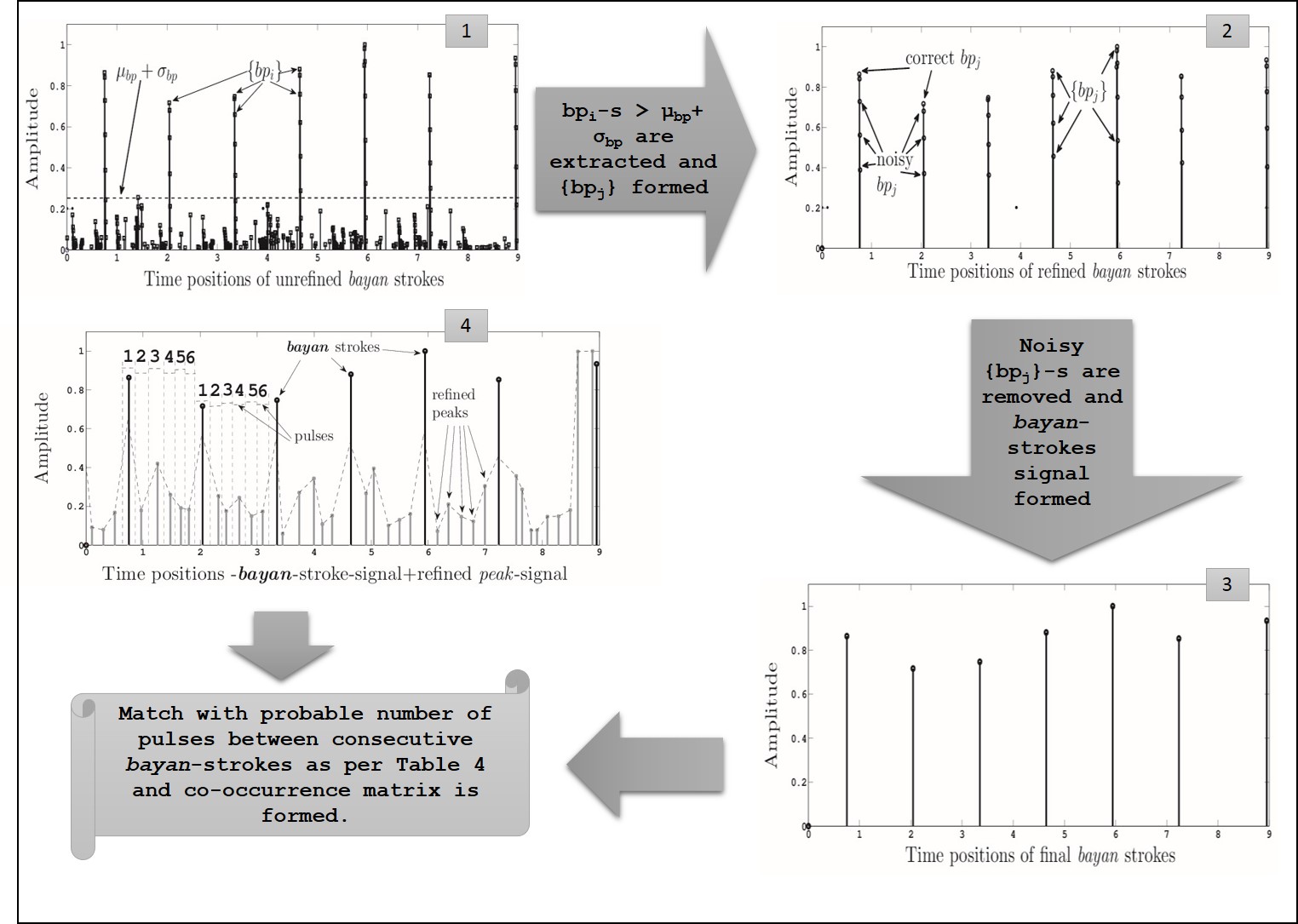}
\caption{Process flow from generation of \textit{bayan}-stroke-signal to the co-occurrence matrix, for a polyphonic composition}
\label{fig_flow}
\end{figure*}

Overall flow of the process starting from generation of \textit{bayan}-stroke-signal to the final co-occurrence matrix is shown in the Figure~\ref{fig_flow}, for a test clip composed in \textit{dadra} \textit{t\={a}la}. The process of generating final \textit{bayan}-stroke-signal[sub-figure 3] is described in the Section~\ref{baya} and final \textit{peak}-signal[sub-figure 4] is described in Section~\ref{allstroke}, Figure~\ref{fig_per}.

\subsection{Separation of \textit{bayan}-stroke-signal}
\label{baya}
In Western music drum is one of the mostly used percussive instruments. Extraction of drum signal is a part of applications like identification of type of drums, re-synthesizing the drum track of a composite music signal.
Existing approaches for drum signal separation are described in Section~\ref{sep} and our method of extracting \textit{bayan}-stroke-signal is described in Section~\ref{our}.
\subsubsection{Drum separation approaches in Western music}
\label{sep}
\begin{enumerate}
\item \textbf{Blind Source Separation method:}
Christian Uhle, Christian Dittmar, Thomas Sporer[~\cite{uhle2003}] proposed a method based on Independent Subspace
Analysis method to separate drum tracks from popular Western music data.
In the work of Hel\'{e}n and Virtanen[~\cite{hel2005}], a method has been proposed for the separation of pitched
musical instruments and drums from polyphonic music. Non-negative Matrix Factorization (NMF) is used to
analyze the spectrogram and thereby to separate the components. 
%For classification of the components Support Vector Machine (SVM) is used.

\item \textbf{Match and adapt method:}
The methodology defines the template(temporal as stated in[~\cite{zil2002}] or spectral as stated in[~\cite{yoshi2004}]),
of drum sound, then searches for similar patterns in the signal. The template is updated and also improved in accordance
with those observations of patterns. This set of methods extracts as well as transcripts drum component. 
%The works presented in[~\cite{zil2002,yoshi2004}] fall under these category.

\item \textbf{Discriminative model}
This approach is built upon a discriminative model between harmonic and drums sounds.
In the work of Gillet and Richard[~\cite{gillet2005}], music signal is split into several frequency
bands, the for each band the signal is decomposed into deterministic and stochastic part. The stochastic part
is used to detect drum events and to re-synthesize a drum track. Possible applications include drum transcription, remixing, and independent processing of the rhythmic and melodic components of music signals. 
Ono et al.[~\cite{ono2008}] have proposed a method that exploits the differences in the spectrograms
of harmonic and percussive components.
\end{enumerate}

\subsubsection{Our approach}
\label{our}
Our approach for separating out \textit{bayan}-stroke-signal falls in the Discriminative model group for separating out harmonic and drums sounds, among the three categories described above. 

To extract the \textit{bayan}-stroke-signal we have used ERB or Equivalent Rectangular Bandwidth filter banks.
The ERB is a measure used in psychoacoustics, which gives an approximation to the bandwidths of the filters 
in human auditory system [~\cite{moore1983}]. 
Alghoniemy and Tewfik[~\cite{algho1999}] have done empirical study of western drums and confirmed that that they could extract the bass drum sequences by filtering the music signal with a narrow bandpass filter. Ranade[~\cite{rana1964}] confirmed the same range($60$-$200$Hz) for the bass drum or \textit{bayan} of Indian \textit{tabl\={a}}. If we take $20$ ERB filter banks to extract different components like voice, \textit{tabl\={a}} and other accompaniments from the polyphonic signal with sampling rate of $44100$Hz, the central frequency of the second bank comes out to be around $130$Hz and the bandwidth of around $60$-$200$Hz. It has been observed from the spectral and wavelet analysis of the different type of \textit{bayan}-\textit{bol}-s described in the Section~\ref{tablaa} and Table~\ref{bol}, that their frequency ranges around the same central frequency and bandwidth. So we have divided the input polyphonic signal sampled at $44100$Hz, into $20$ ERB filter banks and extracted the second bank for constructing the \textit{bayan}-stroke-signal. We have used MIRtoolbox[~\cite{lartillot2007}] to extract this frequency range from ERB filter banks.

As described in the Section~\ref{tala}, most of the \textit{t\={a}la}-s in NIMS start with a highly stressed \textit{t\={a}l\={i}}-\textit{sam-bol} played with \textit{bayan} or \textit{bayan} and \textit{dayan} combined. Moreover \textit{t\={a}l\={i}} \textit{vibh\={a}ga} boundaries are also usually stressed.
Thus extracted \textit{bayan}-stroke-signal would mostly consist of \textit{peak}-s generated from \textit{t\={a}l\={i}}-\textit{sam}-s and \textit{t\={a}l\={i}}-\textit{vibh\={a}ga} boundaries.
There might be presence of other high-strength-\textit{peak}-s generated out of \textit{bol}-s having \textit{bayan} component, other than \textit{t\={a}l\={i}}-\textit{sam} and \textit{t\={a}l\={i}}-\textit{vibh\={a}ga} boundaries, for compositions with slow(20BPM) or very slow(10BPM) tempo. But for popular, semi-classical and filmy North Indian compositions, the tempo is moderate to fast. These compositions mostly do not have the emphasized, high strength \textit{bayan}-\textit{peak}-s other than \textit{t\={a}l\={i}}-\textit{sam}, \textit{t\={a}l\={i}}-\textit{vibh\={a}ga} boundaries in the \textit{bayan}-stroke-signal. 
%For them, the \textit{bayan} \textit{bol}-s other than \textit{sam} or \textit{t\={a}l\={i}}-\textit{vibh\={a}ga} boundaries are much weaker in strength and mostly similar to \textit{dayan} \textit{bol}-s. 
Even if these additional \textit{bol}-s having \textit{bayan}-component, produce \textit{peak}-s in the \textit{bayan}-stroke-signal, their strength is much weaker, compared to \textit{t\={a}l\={i}}-\textit{sam} or \textit{t\={a}l\={i}}-\textit{vibh\={a}ga} boundaries. 

The process below is followed to remove these additional \textit{bol}-s having \textit{bayan} component, from the \textit{bayan}-stroke-signal. 
\begin {itemize}
\item Let $\{bp_i\}$ be the set of \textit{peak}-s in the \textit{bayan}-strokes-signal extracted from a polyphonic song-signal. Please note the Figure~\ref{fig_flow}(1), where ${bp_i}$-s for a particular polyphonic sample of \textit{dadra} \textit{t\={a}la} are shown.
\item We calculate the \textit{mean}$-\mu_{bp}$ and \textit{standard deviation}$-\sigma_{bp}$ for the set $\{bp_i\}$. In the Figure~\ref{fig_flow}(1), the corresponding value of $\mu_{bp}+\sigma_{bp}$ is shown. It should be noted there are lots of noisy \textit{peak}-s with magnitude less than $\mu_{bp}+\sigma_{bp}$.
\item $\{bp_j\}$ is obtained as a subset of $\{bp_i\}$ after selecting the high-strength \textit{bayan}-strokes greater than $\mu_{bp}+\sigma_{bp}$. 
\item $\{bp_j\}$ is the set of strokes mostly containing \textit{t\={a}l\={i}}-\textit{sam}-s and \textit{t\={a}l\={i}}-\textit{vibh\={a}ga} boundaries having \textit{bayan}-component, for a polyphonic composition. In the Figure~\ref{fig_flow}(2), the corresponding time positions of $\{bp_j\}$ are shown.
\item However there would always be some noisy \textit{peak}-s in $\{bp_j\}$, hence a further refinement is done as per the method in Section~\ref{refine} and finally the refined \textit{bayan}-stroke-signal is shown in the Figure~\ref{fig_flow}(3).
\end {itemize}

\subsection{\textbf{Peak-signal:}}
\label{allstroke}
\begin{figure*}[!t]
\centering
\includegraphics[width=1.0\textwidth]{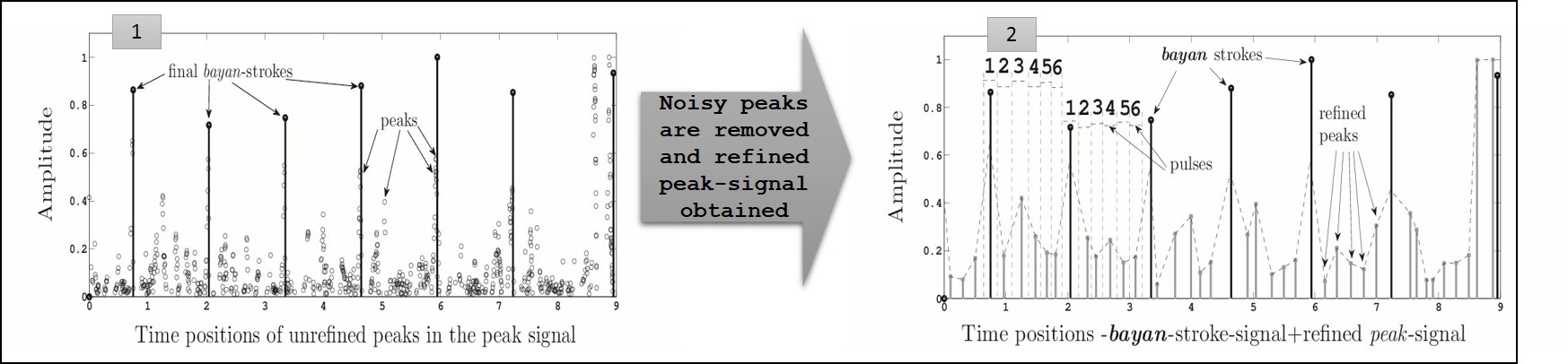}
\caption{Percussive \textit{peak}-s along with \textit{bayan}-strokes in bold line, for a polyphonic clip of \textit{dadra} \textit{t\={a}la}.}
\label{fig_per}
\end{figure*}

From the input polyphonic signal waveform, differential envelope is generated after applying half-wave rectifier. The \textit{peak}-s are extracted from the amplitude envelope of the signal, by calculating the local maxima-s. Local maxima-s are defined as the \textit{peak}-s in the envelope with amplitude higher than their adjoining local minima-s by a default threshold quantity of $d_f \times l_{max}$, where, $l_{max}$ is the maximum amplitude point of the envelope. Here we have taken default minimum value for $d_f$ in the MIRtoolbox[~\cite{lartillot2007}], which would extract almost all the \textit{peak}-s in the input envelope. Each \textit{peak} in the \textit{peak}-signal, is mainly generated out of \textit{tabl\={a}} and other percussive instruments played in the polyphonic composition. These \textit{peak}-s are supposed to be more stressed than other melodic instruments and vocals rendered with comparatively more steady range of energies(without much ups and down, hence unable to produce high-energy \textit{peak}-s).
Figure~\ref{fig_per}(1) shows all the \textit{peak}-s along with the positions of the \textit{bayan}-strokes of the \textit{bayan}-stroke-signal in \textbf{bold}, for the same test clip of the Figure~\ref{fig_flow}. 
The \textit{bayan}-stroke-signal has been generated as per process in Section~\ref{baya}.

\subsection{\textbf{Refinement of \textit{bayan} and \textit{peak}-signal:}}
\label{refine}
There may be multiple percussive instruments and also human voice in a polyphonic composition. There is a tendency to stress more at the \textit{t\={a}l\={i}}-\textit{sam} and \textit{t\={a}l\={i}}-\textit{vibh\={a}ga} boundaries by the performer, while singing along with the \textit{t\={a}la}. Hence for polyphonic compositions, other percussive instruments and human voice also generate \textit{peak}-s in the \textit{bayan}-stroke-signal, coinciding with \textit{bayan}-strokes. 

\textit{Peak}-signal for polyphonic signal also consists of \textit{peak}-s produced by \textit{tabl\={a}}, percussive instruments(if present) and human voice. 
For both \textit{bayan}-stroke and \textit{peak} signals, these \textit{peak}-s should coincide with respect to their positions in $X$-axis or time of their occurrences. But among them the \textit{peak}-s generated out of \textit{tabl\={a}} or the drum instrument here, are usually of higher strength.
Using this theory we go for refinement of both \textit{bayan}-stroke-signal and \textit{peak}-signal to retain the most of the \textit{peak}-s generated from \textit{tabl\={a}}, and discard other kinds of percussive \textit{peak}-s.
It has been observed that most of the popular \textit{hindi} compositions(classical or 
semi-classical) have tempo much less than $600$ beats per minute[~\href{http://raag-hindustani.com/Rhythm.html}{Rhythm Taal}]. Thereby minimum beat interval or gap between consecutive \textit{tabl\={a}} strokes in these compositions, is much more than $60/600 = 0.1$sec.

Hence both the \textit{bayan}-stroke-signal and \textit{peak}-signal are divided 
into $0.1$ sec duration windows along $X$-axis. For each window the \textit{peak} having highest strength is retained as correct \textit{bayan}-stroke(in \textit{bayan}-stroke-signal) or any other valid \textit{tabl\={a}} \textit{peak}(in \textit{peak}-signal), and rest of the \textit{peak}-s in each window is dropped. This way the noisy \textit{peak}-s are removed and final \textit{bayan}-stroke-signal and \textit{peak}-signal are obtained. Figure~\ref{fig_per}(2) shows the final, high-strength and refined \textit{peak}-signal, with the positions of the \textit{bayan}-strokes of the refined \textit{bayan}-stroke-signal in \textbf{bold}. The same final \textit{peak}-signal is referred in Figure~\ref{fig_flow}(4). Figure~\ref{both_d} is the magnified version of Figure~\ref{fig_per}(2),~\ref{fig_flow}(4).

\subsection{\textbf{Analysis based on \textit{tabl\={a}} and \textit{t\={a}la} theory}}
\label{frstpass}

\begin{figure*}[!t]
\centering
\includegraphics[width=1.0\textwidth]{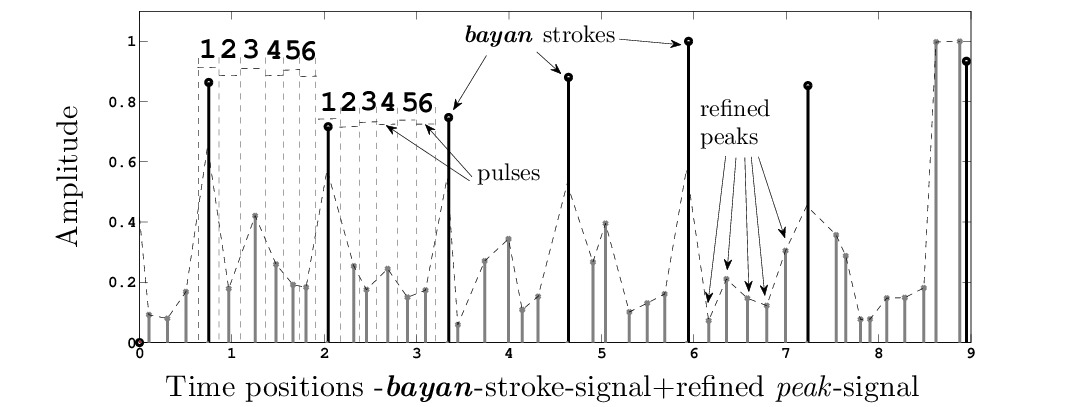}
\caption{Magnified version of Figure~\ref{fig_per}(2) showing details of \textit{peak}-signal.}
\label{both_d}
\end{figure*}

The refined \textit{peak}-signal for the same clip in \textit{dadra} \textit{t\={a}la} is shown in Figure~\ref{both_d}. As per the \textit{thek\={a}} of \textit{dadra} in Table~\ref{theka_table}, apart from the \textit{sam-dha} there is no other \textit{t\={a}l\={i}}-\textit{vibh\={a}ga} boundaries, hence its final \textit{bayan}-stroke-signal should contain these \textit{sam}-s only. 
%susmita282016

\begin{itemize}
\item \textbf{\textit{Pulse:}} Here pulse is defined as the amplitude envelope of a stroke whose \textit{peak} is extracted in the \textit{peak}-signal.
\item \textbf{\textit{Peak:}} It should be noted here that, a \textit{peak} in the refined \textit{peak}-signal(from Section~\ref{refine}), is the highest point of an amplitude envelope formed for a pulse. 
Hence \textit{peak} is actually the mid-point of the pulse duration in seconds along the $X$-axis.
\end{itemize}

For the test clip of \textit{dadra} \textit{t\={a}la}, the \textit{peak}-s and the pulses are elaborated in Figure~\ref{both_d}. Here we can see there are $5$ \textit{peak}-s and $6$ pulses in between two consecutive \textit{bayan} stroke. 
%So for \textit{dadra} \textit{t\={a}la} there should ideally be $5$ peaks and $6$ pulses in between two consecutive \textit{bayan} stroke.  

The method of \textit{t\={a}la} detection based on \textit{tabl\={a}} and \textit{t\={a}la} theory is explained in the following sections. 
%There we can see how the \textit{t\={a}la} of this p
\subsubsection{First level analysis of pulse pattern}
\label{first}
As per the theories explained in Section~\ref{tala} and~\ref{tablaa}, we have extended the Table~\ref{theka_table} and created another Table~\ref{gaprule}. Here the first column describes number of probable pulses in between two \textit{bayan} strokes, as per the theories. For example, for \textit{dadra} \textit{t\={a}la}, number of pulses in between a \textit{dha-bol/sam} of an \textit{\={a}vart} and the \textit{dha-bol/sam} of the next \textit{\={a}vart} should theoretically be $6$ as per the \textit{dadra} \textit{thek\={a}} in the Table~\ref{gaprule}. This $6-6$ pattern of number of pulses should continue along the progression of the composition. 
The third column describes the \textit{thek\={a}}-s corresponding to the pulse pattern, with \textit{t\={a}l\={i}}-\textit{sam}-s and \textit{t\={a}l\={i}}-\textit{vibh\={a}ga}-boundary-\textit{bol}-s in \textbf{bold} as per the theory explained in Section~~\ref{tala} and~\ref{tablaa}. The pipes in \textbf{bold} represent the start of \textit{vibh\={a}ga}-boundaries within single \textit{\={a}vart}. \textit{\={A}vart}-sequences are shown and for each \textit{thek\={a}}, an \textit{\={a}vart} and the starting \textit{bol} of next \textit{\={a}vart} is given, to indicate the progression of \textit{t\={a}la}-s.

It is to be noted that \textit{bhajani} \textit{thek\={a}} has half of its number of strokes as rests. Other
percussive instrument and vocal emphasis would normally generate \textit{peak}-s of moderate strength for the time positions of these strokes in an \textit{\={a}vart}, especially for the genres of our experimental compositions. Hence, here \textit{bhajani} is considered as \textit{t\={a}la} with $16$ pulses/\textit{\={a}vart}.

It should be noted that for \textit{t\={a}la}-s like \textit{bhajani} and \textit{rupak}, there are two sets of probable no of pulses. For example for \textit{rupak} there is both $4-10$ and $10-4$. This is because we are calculating number of pulses for the consecutive \textit{\={a}vart} along the progression of the song. So suppose if we start from \textit{\={a}vart}$^1$, the second \textit{vibh\={a}ga}-boundary-\textit{bol} has \textit{bayan} component and it will generate a \textit{peak} in the \textit{bayan}-stroke-signal. Next \textit{peak} in the \textit{bayan}-stroke-signal would be the third \textit{vibh\={a}ga}-boundary-\textit{bol}. So in between them(I\textit{\textbf{dhin} dhin dha dha}I\textit{\textbf{dhin}}) there would be $3$ \textit{peak}-s and $4$ pulses. Next \textit{peak} in the \textit{bayan}-stroke-signal would be the second \textit{vibh\={a}ga}-boundary-\textit{bol} of the \textit{\={a}vart}$^2$ and evidently there would be $9$ \textit{peak}-s and $10$ pulses between second and third \textit{peak}-s in \textit{bayan}-stroke-signal(\textit{\textbf{dhin} dhin dha dha}$^{2}|$\textit{tun na tun na ti te}I\textit{\textbf{dhin}}). It gives rise to pulse pattern of $4-10$, considering first, second and third \textit{peak}-s in \textit{bayan}-stroke-signal. Now if we move on and consider second, third and fourth \textit{bayan-peak}-s, the detected pulse pattern should be $10-4$, then again $4-10$ and it will go on for the entire progression of the song. So both $4-10$ and $10-4$ would signify \textit{rupak} \textit{t\={a}la} with same set of stressed \textit{bol}-s in the \textit{thek\={a}}.

\begin{table*}[t]
\centering
\caption{Probable number of pulses in between consecutive \textit{bayan}-strokes, corresponding \textit{thek\={a}}-s and the \textit{t\={a}la}-s}
\begin{tabular}{p{0.2\linewidth}p{0.1\linewidth}p{0.6\linewidth}}
\hline
\textbf{Number of probable pulses between consecutive \textit{bayan}-strokes}&\textbf{\textit{T\={a}la}-s}& \textbf{Corresponding \textit{thek\={a}}-s} with \textit{\={a}vart}-sequences\\
\hline
\textbf{6-6}&\textit{dadra}&\textit{$^{1}|$\textbf{dha} dhi na}I\textit{na ti na}$^{2}|$\textit{\textbf{dha}}..\\
\hline
\textbf{8-8}&\textit{kaharba}&\textit{$^{1}|$\textbf{dha} ge na ti}I\textit{na ke dhi na}$^{2}|$\textit{\textbf{dha}}..\\
\hline
\textbf{4-10,10-4}&\textit{rupak}&\textit{$^{1}|$tun na tun na ti te}I\textit{\textbf{dhin} dhin dha dha}I\textit{\textbf{dhin} dhin dha dha}$^{2}|$\textit{tun na}.. \\
\hline
\textbf{14-14}&\textit{rupak}&\textit{$^{1}|$tun na tun na ti te}I\textit{\textbf{dhin} dhin dha dha}I \textit{dhin dhin dha dha}$^{2}|$\textit{tun na}.. \\
\hline
\textbf{3-13,13-3}&\textit{bhajani}&\textit{$^{1}|$\textbf{dhin} $*$ na \textbf{dhin} $*$ dhin na $*$}I\textit{tin $*$ ta tin $*$ tin ta $*$}$^{2}|$\textit{\textbf{dhin} $*$}..\\
\hline
\textbf{16-16}&\textit{bhajani}&\textit{$^{1}|$\textbf{dhin} $*$ na dhin $*$ dhin na $*$}I\textit{tin $*$ ta tin $*$ tin ta $*$}$^{2}|$\textit{\textbf{dhin} } $*$..\\
\hline
\label{gaprule}
\end{tabular}
\end{table*}

\subsubsection{Extended analysis of pulse pattern}
\label{second}
It should be noted that, in \textit{vilambit} compositions there may be additional filler strokes apart from the basic \textit{thek\={a}}, which lead to additional significant \textit{peak}-s in both the \textit{bayan}-stroke-signal and the \textit{peak}-signal. In \textit{druta} compositions often several \textit{thek\={a}} strokes are skipped and only \textit{vibh\={a}ga}-s are stressed.

Table~\ref{gaprule} shows the elementary set of probable pulses in between consecutive \textit{bayan} strokes for clear understanding of the concept. To keep room for variations and improvisations of the \textit{thek\={a}} that are allowed within a specific \textit{t\={a}la}, we have extended this set in our experiment. There we have included all the probable patterns of pulse-counts, by considering the probability of additional \textit{bayan} or \textit{bayan+dayan}-strokes in a \textit{thek\={a}} to be stressed. We are assuming that apart from the mandatory \textit{t\={a}l\={i}}-\textit{sam} and \textit{t\={a}l\={i}}-\textit{vibh\={a}ga} boundaries, any other \textit{bol}-s having \textit{bayan} component may be stressed and produce a \textit{peak} in the \textit{bayan}-stroke-signal.
 
\begin{table*}[t]
\centering
\caption{Probable number of pulses in between consecutive \textit{bayan} strokes in the \textit{thek\={a}} for \textit{dadra} \textit{t\={a}la}}
\begin{tabular}{p{0.3\linewidth}p{0.6\linewidth}}
\hline
\textbf{Number of probable pulses between consecutive \textit{bayan}-strokes} & \textbf{Corresponding \textit{thek\={a}}-s} with \textit{\={a}vart}-sequences\\
\hline
\textbf{6-6}&\textit{$^{1}|$\textbf{dha} dhi na}I\textit{na ti na}$^{2}|$\textit{\textbf{dha}}..\\
\hline
\textbf{1-5,5-1}&\textit{$^{1}|$\textbf{dha} \textbf{dhi} na}I\textit{na ti na}$^{2}|$\textit{\textbf{dha} \textbf{dhi}}..\\
\hline
\label{gaprule1}
\end{tabular}
\end{table*}

Here we have shown all the probabilities(including basic and extended) for \textit{dadra} \textit{t\={a}la} in Table~\ref{gaprule1}, as an example. Here in this \textit{dadra}-\textit{thek\={a}} apart from the \textit{t\={a}l\={i}}-\textit{sam-bol} which is \textit{dha}, of an \textit{\={a}vart}, the very next \textit{bol} is \textit{dhi} also has \textit{bayan}-component. So apart from mandatory \textit{sam} this \textit{bol} can also be stressed and give rise to pulse pattern of $1-5,5-1$.
Similarly for rest of the \textit{t\={a}la}-s, all probable combinations of number of pulses are calculated.

\subsubsection{\textbf{Generation of co-occurrence matrix and detection of \textit{t\={a}la}}}
\label{cooccur}
For each test sample, we have taken the refined version of \textit{bayan}-stroke-signal(generated as per the method in Section~\ref{baya} and then refined as per the method in~\ref{refine}), \textit{peak}-signal(generated as per the method in Section~\ref{allstroke} and refined as per the method in~\ref{refine}) and the co-occurrence matrix is formed and \textit{t\={a}la} is detected as per the following steps. Here co-occurrence matrix displays the distribution of co-occurring pulse-counts along the sequence of the \textit{bayan}-stroke-intervals, in a matrix format The Figure~\ref{fig_flow} shows the overall process flow of generation of co-occurrence matrix from refined \textit{bayan}-stroke-signal and \textit{peak}-signal.
%susmita
\begin{enumerate}
\item We extract the time positions of the \textit{peak}-s of the refined \textit{peak}-signal and the \textit{bayan}-stroke-signal along the $X$-axis or time axis.
\item Then we calculate the count of \textit{peak}-signal-pulses occurring in each of the time intervals formed by consecutive \textit{peak}-s of \textit{bayan}-stroke-signal. Here we denote the series of pulse-counts calculated for a test-sample as $(pc_{1}, pc_{2}, \ldots pc_{k})$, where $k=$(number of \textit{bayan}-strokes$-1$). For example, we can see in the Figure~\ref{both_d}, there are $5$ \textit{peak}-s between two consecutive \textit{bayan}-strokes the number of pulses are $6$ or $pc_{1}=6$. Similarly, we calculate the rest of the $pc_{i}$-s.
\item Then we form a $16X16$ co-occurrence matrix(having $16$ rows and $16$ column/row) and initialize all of its elements with zero. Maximum dimension of the matrix is taken as $16$ because for our test data there can be maximum of $16$ number of pulses between consecutive \textit{bayan}-strokes[Ref Table~\ref{gaprule}].
\item Then we fill up the co-occurrence matrix by occurrence of each pair of pulse-counts between the consecutive intervals in the \textit{bayan}-stroke-signal formed for the whole test sample. We denote each consecutive pair as $pc_{i}$, $pc_{i+1}$, where $pc_{i} \in (1,2, \ldots 16)$ and $pc_{i+1} \in (1,2, \ldots 16)$. For example if the number of pulses between first and the second \textit{peak}-s in the \textit{bayan}-stroke-signal is $4$ and the same between the second and the third is $6$. So $pc_{1}$, $pc_{2}$ becomes $4, 6$ and we add $1$ to the matrix element of fourth row and sixth column, which now becomes $1$ from initialized zero value. Then we check the same between third and fourth \textit{peak} which is suppose $6$, hence $pc_{2}$, $pc_{3}$ becomes $6, 6$ and $1$ is added to the matrix element of sixth row and sixth column, making it $1$ from zero.
\item We traverse the whole \textit{peak}-signal and the \textit{bayan}-stroke-signal and update the matrix. Each cell of the matrix contain the occurrence of a particular pulse count pattern in consecutive intervals in \textit{bayan}-stroke-signal.
\item Finally we extract the row and column index of the cell in the matrix containing the maximum value. This row and and column index is the most occurring pattern of pulse counts in consecutive intervals in \textit{bayan}-stroke-signal. Here this row-column index of the matrix is denoted by [$pcmax_{1}$, $pcmax_{2}$]. 

The co-occurrence matrix for a test sample, is shown in Table~\ref{co-oc} where we can see $10$ as the maximum value in $6^{th}$ row and $6^{th}$ column i.e. [$pcmax_{1}=6$, $pcmax_{2}=6$]. 

\begin{table}[h]
\centering
\caption{Co-occurrence matrix formed for a composition played in \textit{dadra} \textit{t\={a}la}}
%\begin{tabular}{p{0.2\linewidth}p{0.1\linewidth}p{0.6\linewidth}}
%\begin{tabular}{|c|c|c|c|c|c|c|c|c|}
\begin{tabular}{ccccccccc}
\hline
&\textbf{1}&\textbf{2}&\textbf{3}&\textbf{4}&\textbf{5}&\textbf{6}&\textbf{7}&\textbf{...}\\
\hline
\textbf{1}&0&0&0&0&3&0&0&...\\
\hline
\textbf{2}&0&0&0&0&0&0&0&...\\
\hline
\textbf{3}&0&0&0&0&0&0&0&...\\
\hline
\textbf{4}&0&0&0&0&0&0&0&...\\
\hline
\textbf{5}&1&0&0&0&0&0&0&...\\
\hline
\textbf{6}&0&0&0&0&0&\textbf{10}&0&...\\
\hline
\textbf{7}&0&0&0&0&0&0&0&...\\
\hline
\textbf{...}&..&..&..&..&..&..&..&...\\\hline
\label{co-oc}
\end{tabular}
\end{table}

\item Then the [$pcmax_{1}$, $pcmax_{2}$] is matched against the first column of the Table~\ref{gaprule} and also the rules defined in Section~\ref{second}. Accordingly \textit{t\={a}la} is decided from its second column. For the test sample for which the co-occurrence matrix is shown in the Table~\ref{co-oc}, $pcmax_1 = 6$ and $pcmax_2 = 6$ are extracted and it is exactly matched with $6-6$ pattern for number of probable pulses between consecutive \textit{bayan}-strokes, hence it is detected as of \textit{dadra} \textit{t\={a}la}.

\item While matching occurrence pattern of \textit{peak}-s between consecutive \textit{bayan}-\textit{peak}-s, apart from the rules explained in Section~\ref{first} and~\ref{second}, a tolerance of $\pm1$ is considered. For example, if for a test clip we get $6-6$ number of \textit{peak}-s between consecutive \textit{bayan} duration in the \textit{bayan}-stroke-signal, we detect it as of \textit{dadra} \textit{t\={a}la}. But even if we get $5-6,6-5$, then also detect it of \textit{dadra}.
This way we are considering human errors within a narrow range of tolerance.
\end{enumerate}

\subsubsection{\textbf{Detection of tempo}}
\label{tempo}
Tempo or \textit{lay} is detected in terms of pulses per minute as per the method below.
\begin{enumerate}
\item Once we detect [$pcmax_{1}$, $pcmax_{2}$], we get the \textit{t\={a}la} as per the process described in Section~\ref{cooccur}. Then we collect all the consecutive pair of \textit{bayan} durations having $pcmax_{1}$ and $pcmax_{2}$ number of pulses for the whole composition. In the Figure~\ref{both_d}, we can see that for this particular \textit{dadra} clip there are $6$ number of pulses in the intervals between first two \textit{bayan}-strokes and also second, third \textit{bayan}-strokes. 

Suppose these \textit{bayan} durations are denoted by $(bd1_{1}, bd1_{2}, \ldots bd1_{n})$ having $pcmax_{1}$ number of pulses and $(bd2_{1}, bd2_{2}, \ldots bd2_{n})$ having $pcmax_{2}$ number of pulses, where $n$ is the value in the cell of co-occurrence matrix having row index $pcmax_{1}$ and column index $pcmax_{2}$. It basically means that $pcmax_{1}$, $pcmax_{2}$ pair has occurred for $n$ no of times in the co-occurrence matrix and also in the whole test composition.

\item Then all these \textit{bayan} durations are added. We denote that by $bayan_{dur} = \sum_{i=1}^{n} bd1_{i}+\sum_{i=1}^{n} bd2_{i}$. Total number of pulses in these durations are $count_{pulse}=n*(pcmax_{1}+pcmax_{2})$. $bayan_{dur}$ is measured in second.
\item The average duration of a pulse in the composition is calculated as $pulse_{dur}=\frac{bayan_{dur}}{count_{pulse}}$ in second.
\item Then the tempo is calculated as $tempo = \frac{60}{pulse_{dur}}$ in beats per minute.
\end{enumerate}

\section{Experimental details}
\label{exp}
\subsection{Data description}
We have experimented with a number of polyphonic composition of NIMS vocal songs rendered with four popular \textit{thek\={a}}-s of the \textit{t\={a}la}-s, as described in Table~\ref{gaprule}. The test compositions are from \textit{bhajan} or devotional, semi-classical and film-music genres, having \textit{tabl\={a}} and other percussive instruments as accompaniments. The film-music and semi-classical genres are chosen because they mostly maintain similar structures with minimal improvisation and regular tempos as far as rhythm of the compositions is concerned. Hence this test dataset should be suitable for finalizing the elementary layer of the \textit{t\={a}la}-detection system of NIMS.

The \textit{t\={a}la}-s considered are {\it dadra}, {\it kaharba}, {\it rupak}, {\it bhajani}, as most of the songs in above genres are composed in these \textit{t\={a}la}-s. Also we got maximum number of annotated samples of polyphonic songs composed with these \textit{t\={a}la}-s, which helped in rigorous testing and validation process. Also as these \textit{t\={a}la}-s have unique \textit{m\={a}tra}-s and they would produce mostly unique number of \textit{peak}-s between consecutive \textit{bayan}-strokes, so experimenting with sufficient number of test samples composed in these \textit{t\={a}la}-s enabled us to validate the applicability of the initial version of our model. 

The annotated list of \textit{t\={a}la}-wise songs are obtained from [\href{http://www.soundofindia.com/songsearch.asp}{Sound of India} and \href{http://chandrakantha.com/tala_taal/bollywood_songs.html}{FILM SONGS IN VARIOUS TALS}] and also from the albums The Best Of Anup Jalota(Universal Music India Pvt Ltd), Bhanjanjali vol $2$(Venus), Bhajans(Universal Music India Pvt Ltd), Songs Of The Seasons Vol $2$(Shobha Gurtu). The annotations are validated by renowned musician \href{http://www.subhranilsarkar.com}{Subhranil Sarkar}. 
All the song clips are in single channel .wav format sampled at $44100$Hz and are annotated. The clips are of $60$ second duration.
The tempo ranges from \textit{madhya} to \textit{ati-druta} tempo. The tempo of the input samples were calculated by manual tapping by expert musicians and this calculated tempo was assumed to be our benchmark for validation.
The detailed description of the data used is shown in Table~\ref{data}. The tempo is uniformly maintained for the input sound samples of the experiment.
The data reflects variation in terms of genre, types of instruments and voices in the composition,
tempo and \textit{m\={a}tr\={a}} of the compositions.% should establish

\begin{table*}[t]
\caption{Description of data}
\label{data}
\begin{center}
\begin{tabular}{llll} 
\hline
\textit{t\={a}la} & \textit{m\={a}tr\={a}}&Tempo range(in BPM)&No of clips\\
\hline
\textit{dadra}&6&140-320&65\\
\hline
\textit{kaharba}&8&220-400&65\\ 
\hline
\textit{bhajani}&8&300-360&65\\
\hline
\textit{rupak}&7&240-375&65\\
\hline
\end{tabular}
\end{center} 
\end{table*}

\subsection{Results}
Table~\ref{conf} shows the confusion matrix for {t\={a}la} detection. Here the column \textbf{\textit{none}} signifies that the \textit{t\={a}la} of the input clip is NOT detected as any of the input \textit{t\={a}la}-s({\it dadra}, {\it kaharba}, {\it rupak}, {\it bhajani}).
There is an incorrect detection between the pair of {\it kaharba} and {\it bhajani}.  
Few \textit{bhajani} samples have been detected as \textit{kaharba} and vice-versa. 
%As already described in the point 1 of the Section~\ref{scenario}, that 
For a specific \textit{laggi} or variation of \textit{bhajani} \textit{t\={a}la}~\href{http://tablabeats.blogspot.in}{Bhajan taal}, a composition might turn out to be with $8-8$ pulse pattern where $pcmax_{1}=8$, $pcmax_{2}=8$. In this case it would be detected as \textit{kaharba} as per our method. However, this error is not so severe as technically \textit{bhajani} is a variation of \textit{kaharba}[~\cite{david2013}].

Also as per the Table~\ref{gaprule} theoretically $8-8$ pulse pattern is for \textit{kaharba} and $16-16$ is for \textit{bhajani}, i.e. pattern for \textit{bhajani} is exactly twice of \textit{kaharba}. For some rare cases of manual error, while playing \textit{tabl\={a}} in \textit{kaharba}, the \textit{tabl\={a}}-expert might make some \textit{sam}-s less stressed and these \textit{sam}-s might fail to generate \textit{bayan}-\textit{peak}-s in the refined \textit{bayan}-stroke-signal. In these cases \textit{kaharba} might produce $16-16$ pulse pattern and would be detected as \textit{bhajani}. However this is much rare as theoretically for any \textit{tabl\={a}} composition the \textit{t\={a}l\={i}-bol}-\textit{sam} must be stressed.

\begin{table}[h]
\caption{Confusion matrix for \textit{t\={a}la} detection for the clips (all figures in \%)}
\label{conf}
\begin{center}
\begin{tabular}[width=0.75\textwidth]{llllll}
\hline
&\textit{dadra}&\textit{kaharba}&\textit{bhajani}&\textit{rupak}&none\\
\hline
\textit{dadra}&80.85&6.38&06.38&4.26&2.13\\
\hline
\textit{kaharba}&4.17&81.25&8.33&2.08&4.16\\
\hline
\textit{bhajani}&3.57&12.50&78.57&3.57&1.79\\
\hline
\textit{rupak}&3.50&4.50&4.00&86.00&2.00\\
%\textit{rupak}&0.00&0.00&12.00&86.00&2.00\\
\hline
\end{tabular}
\end{center}
\end{table}

Table~\ref{perf-t1} shows the performance of proposed methodology in detecting tempo for different compositions.
In judging the correctness of tempo, a tolerance of $\pm 5\%$ is considered.

\begin{table}
\caption{Performance of tempo detection (all figures in \%)}
\label{perf-t1}
\begin{center}
\begin{tabular}[width=0.75\textwidth]{llllll}
\hline
\textit{t\={a}la}&Correct detection\\
\hline
\textit{dadra}&80.85\\
\hline
\textit{kaharba}&77.08\\
\hline
\textit{bhajani}&80.35\\
\hline
\textit{rupak}&76.00\\
\hline
\end{tabular}
\end{center}
\end{table}

Overall \textit{t\={a}la} and tempo detection performance is shown in Table~\ref{gross}.
It is clear that the proposed methodology performs satisfactorily and that too with wide variety of data.

\begin{table}
\caption{Gross performance of \textit{t\={a}la} and tempo detection (all figures in \%)}
\label{gross}
\begin{center}
\begin{tabular}[width=0.75\textwidth]{ll}
\hline\noalign{\smallskip}
\multicolumn{2}{c}{Average performance}\\
\cline{1-2}
\textit{m\={a}tr\={a}} detection&Tempo detection
\\ 
\noalign{\smallskip}\hline\noalign{\smallskip}
81.59&78.60
\\ 
\noalign{\smallskip}\hline
\end{tabular}
\end{center}
\end{table}

\section{Conclusion}
\label{con}

\begin{enumerate}
\item This paper presents the results of analysis of \textit{tabl\={a}} signal of North Indian polyphonic composition, with the help of new technique by extracting the \textit{bayan} signal.
\item The justification of using \textit{bayan} signal as the guiding signal in case of North Indian polyphonic music and detecting \textit{t\={a}la} using the parameters of NIMS rhythm, has been clearly discussed.
\item A large number of polyphonic music samples from \textit{hindi} vocal songs from \textit{bhajan} or devotional, semi-classical and filmy genres were analyzed for studying the effectiveness of the proposed new method.
\item The experimental result of the present investigation clearly supports the pronounced effectiveness of the proposed technique.
\item We would extend this methodology for studying other features(both stationary and non-stationary) of the all the relevant \textit{t\={a}la}-s of NIMS and designing an automated rhythm-wise categorization system for polyphonic compositions. This system may be used for content-based music retrieval in NIMS. Also a potential tool in the area of music research and training is expected to come out of it.
\end{enumerate}

Limitations of the method is that it can not distinguish between \textit{t\={a}la}-s of same \textit{m{a}tr\={a}}. For example \textit{deepchandi} and \textit{dhamar} \textit{t\={a}la}-s have $14$ number of \textit{m{a}tr\={a}}-s, textit{bol}-s and beats in a cycle. We plan to extend this elementary model of \textit{t\={a}la}-detection system for all the NIMS \textit{t\={a}la}-s, by including other properties like timbral information and nonlinear properties of different kinds of \textit{tabl\={a}} strokes/\textit{bol}-s. We may also attempt to transcript the \textit{t\={a}la}-\textit{bol}-s in a polyphonic composition. This extended version of the model may address the NIMS \textit{t\={a}la}-s which share same \textit{m{a}tr\={a}} and also have variety of \textit{lay}-s. The initial version of the software version of the proposed algorithm is in \href{http://www.dgfoundation.in/taalmaan/}{Talman}, where users can upload relevant .wav files of a polyphonic song played on NIMS \textit{t\={a}la}-s and find out the \textit{t\={a}la} computationally.

\section{Acknowledgement} 
\label{ack}
We thank the \textbf{Department of Higher Education and Rabindra Bharati University, Govt. of West Bengal, India} for logistics support of computational analysis. We also thank renowned musician \href{http://www.subhranilsarkar.com}{Subhranil Sarkar} for helping us to annotate test data, validate test results and Shiraz Ray(\href{http://www.dgfoundation.in}{Deepa Ghosh Research Foundation}) for extending help in editing the manuscript to enhance its understandability.

%\bibliographystyle{splncs}
%\bibliography{tablaNMS}

\end{document}